\documentclass[aps]{revtex4}
\usepackage{graphicx}
\newcommand\nn{\nonumber \\}
\newcommand\bk{{\bf k}}
\newcommand\bp{{\bf p}}
\newcommand\bq{{\bf q}}

\newcommand\br{{\bf r}}
\newcommand\lk{\left( }
\newcommand\rk{\right)}
\newcommand\ltk{\left\{ }
\newcommand\rtk{ \right\} }
\newcommand\ldk{\left[ }
\newcommand\rdk{ \right] }
\newcommand\beq{ \begin{eqnarray} }
\newcommand\eeq{ \end{eqnarray} }
\def\sla{\slash{\!\!\!}} 

\newcommand{\idk}{\int \frac{{\rm d}^3\bk}{(2\pi)^3}}

\newcommand{\idpn}{\int \frac{{\rm d}^3p}{(2\pi)^3}}

\begin{document}
\title{Chiral symmetry and density wave in quark matter}
%\date{\today}
\author{E. Nakano}
\affiliation{Yukawa Institute for Theoretical Physics, Kyoto University, 
             Sakyoku, Kyoto 606-8502, Japan}
\author{T.Tatsumi} 
\affiliation{Department of Physics, Kyoto University, 
                Kyoto 606-8502, Japan}
\begin{abstract}
A density wave in quark matter is discussed at finite temperature, 
which occurs along with the chiral condensation, 
and is described by a dual standing wave 
in scalar and pseudo-scalar condensates on the chiral circle. 
The mechanism is quite similar to that for the spin density wave 
suggested by Overhauser, and entirely reflects many-body effects. 
It is found 
%%%%%%%%%%%%second-revised-1%%%%%%%%%%%%%%%%%%%%%%%%%%%%%%%%%%%%%%
within a mean-field approximation for NJL model 
%%%%%%%%%%%%%%%%%%%%%%%%%%%%%%%%%%%%%%%%%%%%%%%%%%%%%%%%%%%%%%%%%%
that the chiral condensed phase with the density wave 
develops at a high-density region  
just outside the usual chiral-transition line in phase diagram. 
A magnetic property of the density wave is also elucidated. 
\end{abstract}
%PACS: 11.30.Rd; 75.10.-b; 26.60.+c\\
%Keywords: quark matter;chiral symmetry; chiral density wave 
\maketitle 

\section{Introduction} 

In a recent decade 
the condensed matter physics of QCD has been an exciting area in nuclear physics. 
In particular, 
its phase structure at finite-density and 
relatively low-temperature region is studied actively, 
since it was suggested that the color superconductivity (CSC) 
involved observable consequence of quark matter 
due to the large magnitude of its gap energy over a few hundred MeV 
\cite{csc0,csc1,csc2}. 
Although 
QCD at finite densities has important implications 
for physics in other fields, 
e.g., the core of the compact stars and their evolution \cite{nutr1},  
it remains in not understood adequately. 

Whereas the quark Cooper-pair ({\it p-p}) condensations attracts much interest, 
particle-hole ({\it p-h}) condensations,  
which are related to the chiral condensation \cite{csc1,csc2} 
or ferromagnetism in quark matter \cite{tat,nak,ferr1}, 
have also been studied, and their interplay with CSC has been discussed.  
Since the Cooper instability on the Fermi surface occurs 
for arbitrarily weak interaction, 
the {\it p-p} condensation should dominate 
at asymptotically free high-density limit.   
While on the other hand, 
there exists a critical strength of the interaction in the {\it p-h} channel 
for its condensation. 

At moderate densities, however, where the interaction is strong enough,  
{\it p-p} and {\it p-h} condensations are competitive, and  
various types of the {\it p-h} condensations are proposed \cite{der1, der2}
in which the {\it p-h} pairs in scalar or tensor channels 
have a finite total momentum indicating standing waves.  
Instability for the density wave in quark matter was first discussed 
by Deryagin {\it et al.} \cite{der0} at asymptotically high densities 
where the interaction was very weak, 
and they concluded that the density-wave instability prevailed over the Cooper's one 
in the large $N_c$ (the number of colors) limit 
due to the dynamical suppression of colored {\it p-p} pairings. 
 
In general, 
the density waves are favored in one-dimensional (1-D) systems,  
and have the wavenumber $Q=2k_F$ according to the Peierls instability 
\cite{kagoshima,peiel1}, 
e.g., charge density waves in quasi-1-D metals \cite{peiel2}. 
The essence of its mechanism is the nesting of Fermi surfaces 
and the level-repulsion (crossing) of single particle spectra 
due to the {\it p-h} interaction with the finite total wavenumber. 
Thus the low dimensionality has a essential role 
to make the density-wave states stable. 
In higher dimensional systems, however, the transition occurs  
provided the interaction of a corresponding ({\it p-h}) channel is strong enough. 
For the 3-D electron gas, 
it was shown by Overhauser \cite{ove,sdw1} that the paramagnetic state was unstable 
with respect to formation of a static spin density wave (SDW), 
in which spectra of up- and down-spin states deform 
to bring about a level-crossing 
due to the Fock exchange interactions, 
while the wavenumber does not precisely coincide with $2k_F$ 
because of incomplete nesting in the higher dimension. 

In the recent paper \cite{LettCDCW},  
we suggested a density wave in quark matter at moderate densities   
in analogy with SDW mentioned above. 
It occurs along with the chiral condensation 
and is represented by 
a dual standing wave in scalar and pseudo-scalar condensates 
(we have called it `dual chiral-density wave', DCDW). 
The DCDW has different features in comparison with 
the previously discussed chiral-density waves \cite{der0,der1,der2}, 
and emerges at a moderate density region $\rho_B/\rho_0 \simeq 3-6$ 
(where $\rho_0=0.16 {\rm fm}^{-3}$, the normal nuclear matter density). 

In this paper 
we would like to further discuss DCDW and figure out the mechanism in detail.
We also present a phase diagram on the density-temperature plane.  
In Sec. II 
we start by introducing the order parameters for the dual chiral-density wave 
and show the nature of the ground state on the analogy of the spin density wave:  
a level-crossing of single-particle spectrum in the left- and right-handed quarks. 
Sec. III is devoted to concrete calculations 
by use of Nambu-Jona-Lasinio (NJL) model \cite{nam}  
to present a phase diagram at finite density and temperature, 
in the case of 2 flavors and 3 colors. 
In Sec. IV we summarize and give some comments on outlooks.

%%%%%%%%%%%%%%%%%%%%%%%%%%%%%%%%%%%%%%%%%%%%%%%%%%%%%%%%%%%%%%%%%
\section{Nature of dual chiral-density wave}
In the pioneering studies of chiral density waves \cite{der0,der1,der2},   
a spatial modulation in the chiral condensation was considered; 
the scalar condensation with a wavenumber vector ${\bf q}$ occurs, 
$\langle\bar{\psi}\psi\rangle \propto \cos({\bf q}\cdot {\bf r})$.  
In this section, 
we consider a directional modulation with respect to the chiral rotation, 
since the directional excitation modes should be lower than the radial ones 
in the spontaneously symmetry-broken phases. 
We propose the dual chiral-density wave (DCDW)
in scalar and pseudo-scalar condensation, 
\begin{eqnarray}
\langle\bar{\psi} \psi \rangle &=& \Delta \cos({\bf q}\cdot {\bf r}),  \nonumber \\
\langle\bar{\psi} i \gamma_5 \psi  \rangle  &=& \Delta \sin({\bf q}\cdot {\bf r}), 
\label{dcdw1}
\end{eqnarray}
where the amplitude $\Delta$ corresponds 
to the magnitude of the chiral condensation; 
$\langle\bar{\psi} \psi \rangle^2
+\langle\bar{\psi} i \gamma_5 \psi  \rangle^2  = \Delta^2$. 

\begin{figure}
\begin{center}
\includegraphics[height=4cm]{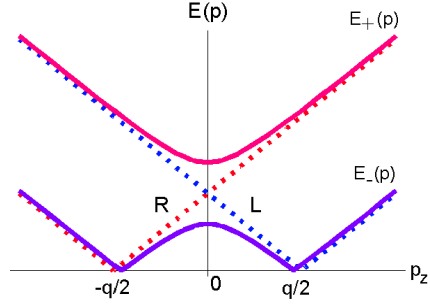}
\end{center}
\caption{Two branches of the single-particle spectrum (\ref{SPS1}) are 
plotted along $p_z$ axis at $p_x=p_y=0$ where ${\bf q}//\hat{z}$. 
Dashed (Solid) lines correspond to $M=0$ ($M\neq 0$). }
\label{levelcross1}
\end{figure}

We give the single-particle spectrum,  
in the presence of the density wave (\ref{dcdw1}) as an external field.  
The Lagrangian density of the system in the chiral limit then reads 
\begin{eqnarray}
%\int {\rm d}^3r 
{\cal L} &=& 
%\int {\rm d}^3r ~
\bar{\psi}(r) 
\ldk 
i\sla{\partial} 
+ 2G\Delta \ltk \cos({\bf q}\cdot {\bf r})
  +i\gamma_5\sin({\bf q}\cdot {\bf r}) \rtk  
\rdk \psi(r), 
\label{Lag1}
\end{eqnarray}
where $2G$ is a coupling constant between the quark and DCDW.   
%$\sigma^\mu=(1,{\bf \sigma})$, $\bar{\sigma}^\mu=(1,-{\bf \sigma})$.   
In the chiral representation,   
it becomes clear how the density wave affects the quark fields:   
the Lagrangian %eq.~(\ref{eqLag1}) 
shows that the density wave connects left- and right-handed particles 
in pairs with the total momentum ${\bf q}$, 
\begin{eqnarray}
\int {\rm d}^3r{\cal L}  &=& \int \frac{{\rm d}^3p}{(2\pi)^3} 
\left( \begin{array}{c}
\psi_L(p-q/2) \\
\psi_R(p+q/2)
\end{array} \right)^\dagger
\left( \begin{array}{cc}
\bar{\sigma}_\mu (p-q/2)^\mu & -M  \\
  -M & \sigma_\mu (p+q/2)^\mu
\end{array}  \right)
\left( 
\begin{array}{c}
\psi_L(p-q/2) \\
\psi_R(p+q/2)
\end{array}  \right),  
\label{eqLag1}
\end{eqnarray}   
where $q^\mu=(0,{\bf q})$, and $M$($\equiv -2G\Delta$) corresponds to the dynamical mass 
if the density wave is generated by quark interactions as the mean-field.   
The single-particle (quasi-particle) spectrum is obtained from the poles of the propagator: 
\begin{eqnarray}
\det{\left[ \bar{\sigma}_\mu (p-q/2)^\mu \sigma_\mu (p+q/2)^\mu-M^2 \right]}=0. 
\end{eqnarray}
From the above equation  
we found the spectrum: positive and negative energies, 
$E_\pm(\bp)$ and $-E_\pm(\bp)$, 
\begin{eqnarray}
E_{\pm}(\bp)= \sqrt{E_0(\bp)^{2}+|{\bf q}|^2/4\pm 
\sqrt{({\bf p}\cdot{\bf q})^2+M^{2}|{\bf q}|^2}},~~~E_0(p)=(M^2+|{\bf p}|^2)^{1/2} 
\label{SPS1}. 
\end{eqnarray}
Because of the finite ${\bf q}$ and $\Delta$,   
the spectrum is deformed and split into two branches denoted by $E_\pm(\bp)$, 
as shown in Fig.~\ref{levelcross1} 
where the direction of ${\bf q}$ is taken parallel to the $z$ axis. 
The two branches exhibit a level-crossing between the left- and right-handed particles.  
%------------------------------------------------------
%------------------------------------------------------   

%%%%%%%%%%%%%%%%%%%%%%%%%%%%%%%%%%%%%%%%%%%%%
For $\bq=(0,0,q)$, 
the corresponding eigen spinors are given, in the chiral representation, by 
\begin{eqnarray}
u_\pm(\bp) &\equiv& \mathcal{N_\pm}
\lk
\begin{array}{c} 
    \psi_L^{(\pm)}(\bp-\bq/2) \\
    \psi_R^{(\pm)}(\bp+\bq/2)
  \end{array} \rk ~~\mbox{for positive energy } E_\pm(\bp), 
\label{sp1} \\ 
v_\pm(\bp)&\equiv& u_\pm(\bp)|_{E_\pm(\bp)\rightarrow -E_\pm(\bp)}
~~\mbox{for negative energy} -E_\pm(\bp), 
\label{sp2} 
\end{eqnarray}
where $\mathcal{N_\pm}$ is a normalization factor 
for $u_\pm^\dagger u_\pm=v_\pm^\dagger v_\pm=1$, 
and 
\beq
\psi_R^{(\pm)}(\bp+\bq/2)
&=& 
\lk
\begin{array}{c} 
    \frac{\pm\sqrt{p_z^2+M^2}+q/2+E_\pm(\bp)}{p_x+ip_y} \\
    1
  \end{array} \rk, \\
\psi_L^{(\pm)}(\bp-\bq/2)
&=&
 \frac{M}{p_z\pm\sqrt{p_z^2+M^2}}~\sigma_3~\psi_R^{(\pm)}(\bp+\bq/2).  
\eeq
Here 
it should be noted that the original spinor in eq.~(\ref{Lag1}), $\psi(r)$, 
is represented as a plane wave expansion for the eigen spinors in eqs.~(\ref{sp1})-(\ref{sp2}): 
\beq
\psi(r)=e^{-i\gamma_5 \bq \cdot {\bf r}/2} \sum_{s=\pm} 
\int \frac{{\rm d}^3p}{(2\pi)^3} \ltk a_s(\bp) u_s(\bp) + b_s(\bp) v_s(\bp) \rtk e^{-ip \cdot r}
\equiv 
e^{-i\gamma_5 \bq \cdot {\bf r}/2} \int \frac{{\rm d}^3p}{(2\pi)^3} \tilde{\psi}(p) e^{-ip \cdot r}, 
\label{sp3} 
\eeq
where $a_\pm(\bp)$ ($b_\pm(\bp)$) is annihilation operator for the positive (negative) quasi-particle state.   
The factor, $\exp(-i\gamma_5 \bq \cdot {\bf r}/2)$, 
comes from the momentum shifts $\mp q/2$ 
in the left- and right-handed particles in eq.~(\ref{eqLag1}) due to the presence of DCDW, 
and thus reflects a nature of the spatially chiral-rotated ground state. 
The factor corresponds also to a kind of Weinberg transformation,  
which changes the system from spatially modulated one to a uniform one \cite{LettCDCW,wei}.     

%%%%%%%%%%%%%%%%%%%%%%%%%%%%%%%%%%%%%%%%
In the previous section, 
we mentioned that in general, density waves should be favored in 1-D systems. 
If we assume a quasi-1-D system along the direction of the $z$ axis, 
suppressing the radial ($x$-$y$) degrees of freedom,    
a gap ($\simeq 2M$) opens just above the Fermi surface, 
provided that $q$ is taken to be $2 k_F$ 
($k_F$: the Fermi momentum for free quarks), 
as illustrated in Fig.~\ref{levelcross1}.  
In this case,  
only the lower branch is occupied,  
and the total energy lowers 
for formation of the density wave with wavenumber $2k_F$ 
due to the nesting effect \cite{peiel1,peiel2}. 
In the uniform 3-D system 
we discuss in this article, however, 
the wavenumber dynamically depends on 
the balance between the kinetic and interaction energies, 
and becomes smaller than $2k_F$:  
the spatial modulation due to DCDW makes the kinetic-energy loss, and 
the energy gain is generated by the deformation of the single-particle spectrum 
which originates from the {\it p-h} interaction. 
This situation is essentially the same as SDW 
in 3-D electron system discussed by Overhauser \cite{ove}. 
In the next section 
we demonstrate the actual manifestation of DCDW 
by taking a definite model.

\section{Application to the NJL model}
Since DCDW defined by eq.~(\ref{dcdw1}) 
is associated with the chiral condensation, 
we consider the moderate density region 
where nonperturbative phenomena are expected to remain even in quark matter. 
We here employ the NJL Lagrangian with $N_f=2$ flavors and $N_c=3$ colors 
\cite{nam,kle} to describe such a situation, 
\begin{eqnarray}
{\cal L}_{NJL}
=\bar\psi(i\sla{\partial}-m_c)\psi+G[(\bar\psi\psi)^2+
(\bar\psi i\gamma_5\mbox{\boldmath$\tau$}\psi)^2], 
\label{njl}
\end{eqnarray}
where {\boldmath$\tau$} is isospin matrix, and
$m_c$ is the current mass, $m_c\simeq 5$MeV. 
 
We assume the mean-fields in the direct (Hartree) channels,
\begin{eqnarray}
\langle\bar\psi\psi\rangle&=&\Delta\cos({\bf q}\cdot {\bf r}) \nonumber \\
\langle\bar\psi i\gamma_5\tau_3\psi\rangle&=&\Delta\sin({\bf q}\cdot {\bf r}), 
\label{chiral1}
\end{eqnarray}
where we fix the isospin direction to $\tau_3$ 
because of degeneracy on the isospin hyper sphere; 
the other mean-fields vanish consistently, 
$\langle\bar\psi i\gamma_5\tau_1\psi\rangle =
\langle\bar\psi i\gamma_5\tau_2\psi\rangle=0$. 
%%%%%%%%%%Newly-Added-0%%%%%%%%%%%%%%%%%%%%%%%%%%%%%%%%%%
In other words, 
it is assumed that DCDW is a charge eigen state, and 
there is no amplitude which mixes states with different charges 
%%%%%%%%%%Newly-Added-0%%%%%%%%%%%%%%%%%%%%%%%%%%%%%%%%%%
%
\footnote{Note that this configuration is not unique, 
but in general more complicated cases can be considered, 
e.g., multi-standing waves such as 
$\langle\bar\psi \psi\rangle=\Delta\cos({\bf q}\cdot {\bf r})$ and 
$\langle\bar\psi i\gamma_5\tau_i\psi\rangle=\Delta\sin({\bf q}\cdot {\bf r}) e_i$, 
where 
$e_i=\{\sin({\bf p_1}\cdot {\bf r}) \cos({\bf p_2}\cdot {\bf r}), 
\sin({\bf p_1}\cdot {\bf r}) \sin({\bf p_2}\cdot {\bf r}), 
\cos({\bf p_1}\cdot {\bf r})\}$, 
and ${\bf q}$, ${\bf p_{1,2}}$ are independent wavenumber vectors. 
This configuration is also in the chiral circle: 
$\langle\bar\psi \psi\rangle^2 
+\langle\bar\psi i\gamma_5\tau_i\psi\rangle^2=\Delta^2$.}. 
%
%%%%%%%%%%Newly-Added-1%%%%%%%%%%%%%%%%%%%%%%%%%%%%%%%%%%
As for the Fock exchange terms, 
we have briefly examined them in Appendix~\ref{fock}, and 
shown that the tensor exchange terms might affect DCDW implicitly 
through the deformation of one-particle spectrum. 
In the present study, however, 
we will treat only the direct terms 
since the exchange terms correspond to pure quantum processes 
and have less contribution than the direct ones. 
%%%%%%%%%%Newly-Added-1%%%%%%%%%%%%%%%%%%%%%%%%%%%%%%%%%%
% 
It is interesting that the configuration (\ref{chiral1}) 
is similar to the pion condensation in high-density nuclear 
matter within the $\sigma$ model,  
suggested by Dautry and Nyman \cite{dau}, where $\sigma$ and $\pi^0$ 
meson condensates take the same form as eq.~(\ref{chiral1}). 
It may implies a kind of quark-hadron continuity \cite{QHC1}. 

Within the mean-field approximation, 
the effective Lagrangian becomes 
\begin{eqnarray}
{\cal L}_{MF}=\bar{\psi} \left[i\sla{\partial} +\mu \gamma_0
-M\{\cos({\bf q}\cdot {\bf r})
  +i\gamma_5 \tau_3 \sin({\bf q}\cdot {\bf r}) \} \right]\psi -\frac{M^2}{4G},
\label{effl}
\end{eqnarray}
where we have introduced the chemical potential $\mu$, 
and taken the chiral limit ($m_c=0$) assuming $\mu >> m_c$.   
Since only the difference between u- and d- quark is the sign of the wavenumber vector 
(${\bf q}\leftrightarrow -{\bf q}$) due to the isospin matrix $\tau_3$, 
the single-quark spectrum takes the same form as in eq.~(\ref{SPS1}). 
Thus we need not distinguish two flavors in the energy spectrum, 
though the eigen spinors depend on the sign of ${\bf q}$.

\begin{figure}
\begin{center}
\includegraphics[height=4.2cm]{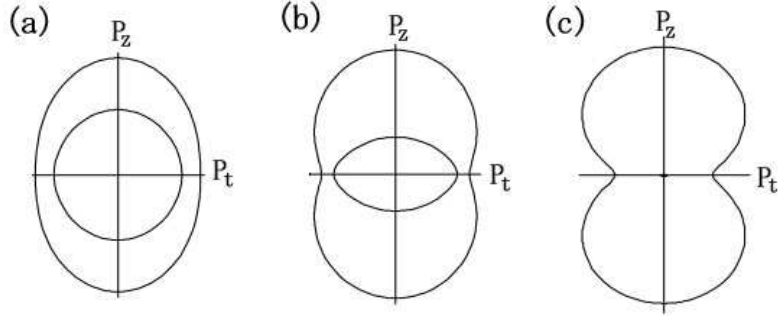}
\end{center}
\caption{Fermi surfaces of the spectrum (\ref{SPS1})
for ${\bf q}//\hat{z}$. 
$p_t=\sqrt{p_x^2+p_y^2}$. ~
The outer (inner) closing line corresponds to $E_-(p)$ ($E_+(p)$). 
(a) for $M\ge q/2$; the minimum of $E_-(p)$ is at the origin. 
(b) for $M\le q/2$; 
there are two minima at points 
$(p_t,p_z)=\left(0, \pm\sqrt{(q/2)^2-M^2}\right)$.  
(c) for $M\le q/2$ and $\mu=q/2+M$; the minor Fermi sea vanishes.}
\label{FS1}
\end{figure}

Hereafter, 
we take the direction of the wavenumber vector parallel to the $z$ axis, 
${\bf q}=(0,0,q)$ without loss of generality, and show the Fermi surface 
for various values of $\mu$, $M$, and $q$ in Fig.~\ref{FS1}. 
%------------------------------------------------------
%------------------------------------------------------   
%zero at ${\bf p}=(0,0,\sqrt{q^2/4-m^2})$ 
%as a function of $|{\bf p}|$ in case of $m<q/2$.
%figure2

In the density-wave state, 
all the energy levels below the chemical potential are occupied 
in the deformed spectrum. 
Accordingly the thermodynamic potential density at zero temperature becomes 
\begin{eqnarray}
\Omega_{\rm tot}&=&N_f N_c \int\frac{{\rm d}^3p}{(2\pi)^3} \sum_{s=\pm}
\left[\left\{ E_s(p)-\mu \right\} \theta\left(\mu-E_s(p) \right)-E_s(p) \right] 
+\frac{M^2}{4G} \nonumber \\
&\equiv&\Omega_{\rm fer}+\Omega_{\rm vac}+\frac{M^2}{4G}, 
\label{therm}
\end{eqnarray}
where $\Omega_{\rm fer}$ ($\Omega_{\rm vac}$) 
denotes the Fermi-sea (Dirac-sea) contribution. 
We can see that the finite wavenumber effect enters 
only through the deformation of the energy spectrum, 
and gives non-trivial contributions 
to the behaviour of the chiral condensation or the dynamical mass $M$. 
The energy gap between the two branches is generated by the dynamical mass 
which comes mainly from the Dirac sea, and  
the energy gain due to the density wave (to a finite $q$) comes 
essentially from the Fermi sea 
which is responsible for the finite baryon-number density, 
thus the DCDW is produced cooperatively by the Dirac and Fermi seas.     

Since the NJL model is unrenormalizable, 
we need some regularization procedure 
to evaluate the negative-energy contribution $\Omega_{\rm vac}$. 
Because of the spectrum anisotropy  
we cannot apply the momentum cut-off regularization scheme.  
Instead we adopt the proper-time regularization (PTR) scheme \cite{sch}. 
We show the result (the derivation is detailed in Appendix~\ref{PTR1}), 
\begin{eqnarray}
\Omega_{\rm vac}=\frac{\gamma}{8\pi^{3/2}}\int_{1/\Lambda^2}^\infty
\frac{{\rm d}\tau}{\tau^{5/2}}
\int^\infty_{-\infty}\frac{{\rm d}p_z}{2\pi}\left[
e^{-(\sqrt{p_z^2+m^2}+q/2)^2\tau}
+e^{-(\sqrt{p_z^2+m^2}-q/2)^2\tau}\right]-\Omega_{{\rm ref}},
\label{omevac}
\end{eqnarray}
where $\Lambda$ is the cut-off parameter, 
and we subtracted an irrelevant constant $\Omega_{{\rm ref}}$ 
in the derivation.  
All the physical quantities should be taken to be smaller than the scale $\Lambda$ 
in the following calculations.

\subsection{Phase transition at zero temperature}

\begin{figure}[h] 
\begin{center}
\includegraphics[height=5cm]{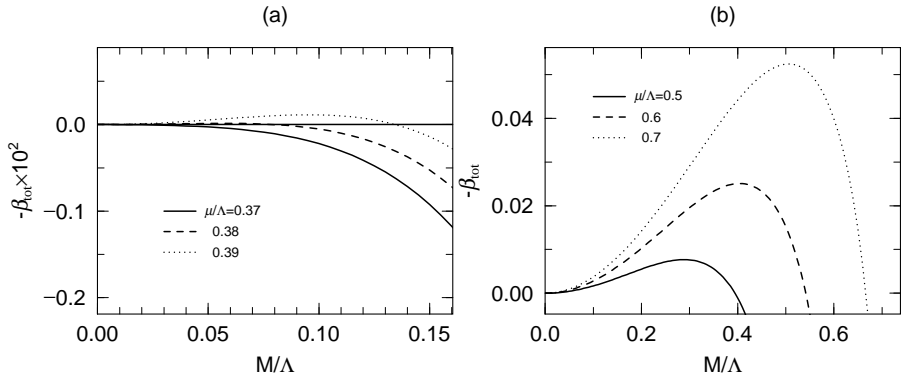}
\end{center}
\caption{The dynamical-mass $M$ dependence of 
the coefficient $\beta_{tot}$ 
for various values of the chemical potential, 
(a) for $\mu/\Lambda=0.37-0.39$, (b) for $\mu/\Lambda=0.5-0.7$. }
\label{betatot}
\end{figure}

To investigate threshold density for formation of the density wave at $T=0$, 
we expand the potential (\ref{therm}) up to the second order in $q$, 
and examine the sign of its coefficient, 
%\footnote{Note that the higher order terms of $q$ diverge 
%as $\Lambda\rightarrow\infty$, and thus are discarded \cite{sug}.}
\begin{eqnarray}
\Omega_{\rm tot}&=& \Omega_{\rm tot}^0
+\frac{1}{2}(\beta_{\rm fer}+\beta_{\rm vac})q^2+O(q^4) \\
\beta_{\rm fer}&\equiv &
\frac{\partial^2 \Omega_{\rm fer}}{\partial q^2}|_{q\rightarrow0}
= -N_f N_c \frac{M^2}{\pi^2}H(\mu/M) \label{betaF} \\
\beta_{\rm vac}&\equiv &
\frac{\partial^2 \Omega_{\rm vac}}{\partial q^2}|_{q\rightarrow0} 
=N_f N_c \frac{\Lambda^2}{2\pi^2}J(M^2/\Lambda^2), 
\label{betaV} 
\end{eqnarray}
where $J(x)=x\int_x^\infty {\rm d}\tau \exp(-\tau)/\tau $, 
and $H(x)={\rm ln}(x+\sqrt{x^2-1})$.
The coefficient of the second-order term in $\Omega_{\rm fer}$ 
is always positive for finite dynamical mass $M\neq 0$, 
while the counterpart of $\Omega_{\rm vac}$ is negative, 
indicating that the Dirac sea is stiff 
against the formation of the density wave.  
In contrast, the Fermi sea favors it as mentioned in the previous section. 
The total coefficient, 
$\beta_{\rm tot} \equiv \beta_{\rm fer}+\beta_{\rm vac}$, 
depends on the dynamical mass and the chemical potential for fixed $\Lambda$, 
as shown in Fig.~\ref{betatot}. 
%figure3
%------------------------------------------------------
%------------------------------------------------------   
%
For larger values of the chemical potential in Fig.~\ref{betatot}(b), 
$\beta_{{\rm tot}}$ becomes negative and reaches its maximum at a finite $M$. 
As for the small chemical potential $\mu/\Lambda<0.38$ in Fig.~\ref{betatot}(a), 
$\beta_{\rm tot}$ never become negative for any value of $M$. 
It leads to a rough estimation of the critical coupling constant, 
$G\Lambda^2\simeq 4.63$, which is the value to occur the usual chiral condensation ($q=0$) 
at $\mu/\Lambda=0.38$ in the PTR scheme. 
%Nevertheless, the density wave needs only 
%about a half values of magnitude of the ordinary chiral condensation ($q=0$), 
%as shown later. 

The magnitudes of $M$ and $q$ are obtained from the minimum 
of the potential (\ref{therm}) at $T=0$, 
and their values satisfy the stationary conditions, 
$\partial \Omega_{tot}/\partial M=\partial \Omega_{tot}/\partial q=0$.  

\begin{figure}[h] 
\vspace*{-0cm}
\begin{center}
\includegraphics[height=14.5cm, angle=-90]{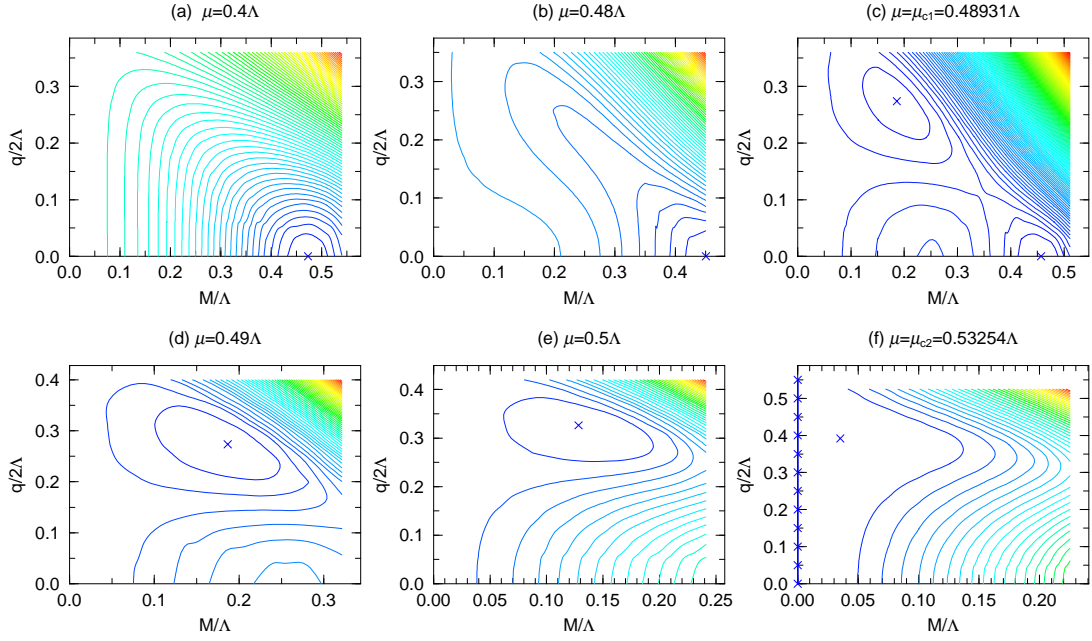}
\end{center}
\caption{Contours of $\Omega_{tot}$ at $T=0$ are shown in $M$-$q$ plane 
as the chemical potential increases, (a) $\rightarrow$ (f).  
The cross in each figure denotes the absolute minimum.}
\label{cp1}
\end{figure}

Fig.~\ref{cp1} shows contours of $\Omega_{\rm tot}$ in $M$-$q$ plane  
for a given chemical potential,    
where the parameters are set as $G\Lambda^2=6$ and $\Lambda=850$ MeV, 
which are not far from those for the vacuum ($\mu=0$) \cite{kle}. 
%figure4
%------------------------------------------------------
%------------------------------------------------------   
The crossed points denote the absolute minima. 
There appear two critical chemical potentials $\mu=\mu_{c1}, \mu_{c2}$:    
for the lower densities (Fig.~\ref{cp1}(a)-(b)) 
the absolute minimum lies at the point $(M\neq 0, q=0)$ 
indicating a finite chiral condensation. 
At $\mu=\mu_{c1}$ (Fig.~\ref{cp1}(c)) 
the potential has the two absolute minima at $(M\neq 0, q=0)$ and $(M\neq 0, q\neq 0)$, 
showing the first-order transition to the DCDW phase, 
which is stable for $\mu_{c1}<\mu<\mu_{c2}$ (Fig.~\ref{cp1}(d)-(e)). 
At $\mu=\mu_{c2}$ (Fig.~\ref{cp1}(f))    
any point on the line $M=0$ and a point $(M\neq 0, q\neq 0)$ become minimum, 
and thereby 
the system undergoes the first-order transition to the chiral-symmetric phase 
which is stable for $\mu>\mu_{c2}$.    

\begin{figure}[h] 
\begin{center}
\includegraphics[height=6cm]{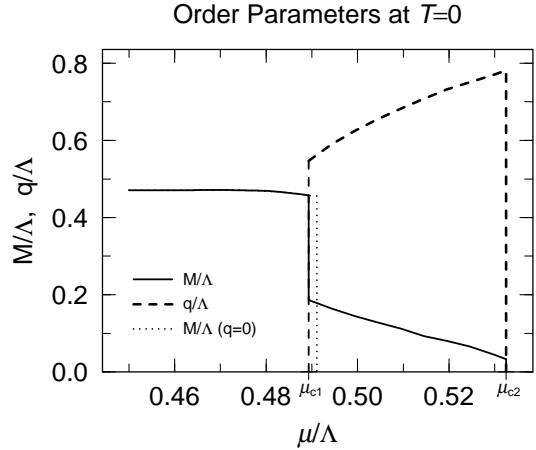}
\end{center}
\caption{Wavenumber $q$ and dynamical mass $M$ are plotted 
as functions of the chemical potential at $T=0$. 
Solid (dotted) line for $M$ with (without) the density wave, 
and dashed line for $q$.}
\label{op1}
\end{figure}

\begin{figure}[h]
\begin{center}
\includegraphics[height=6cm]{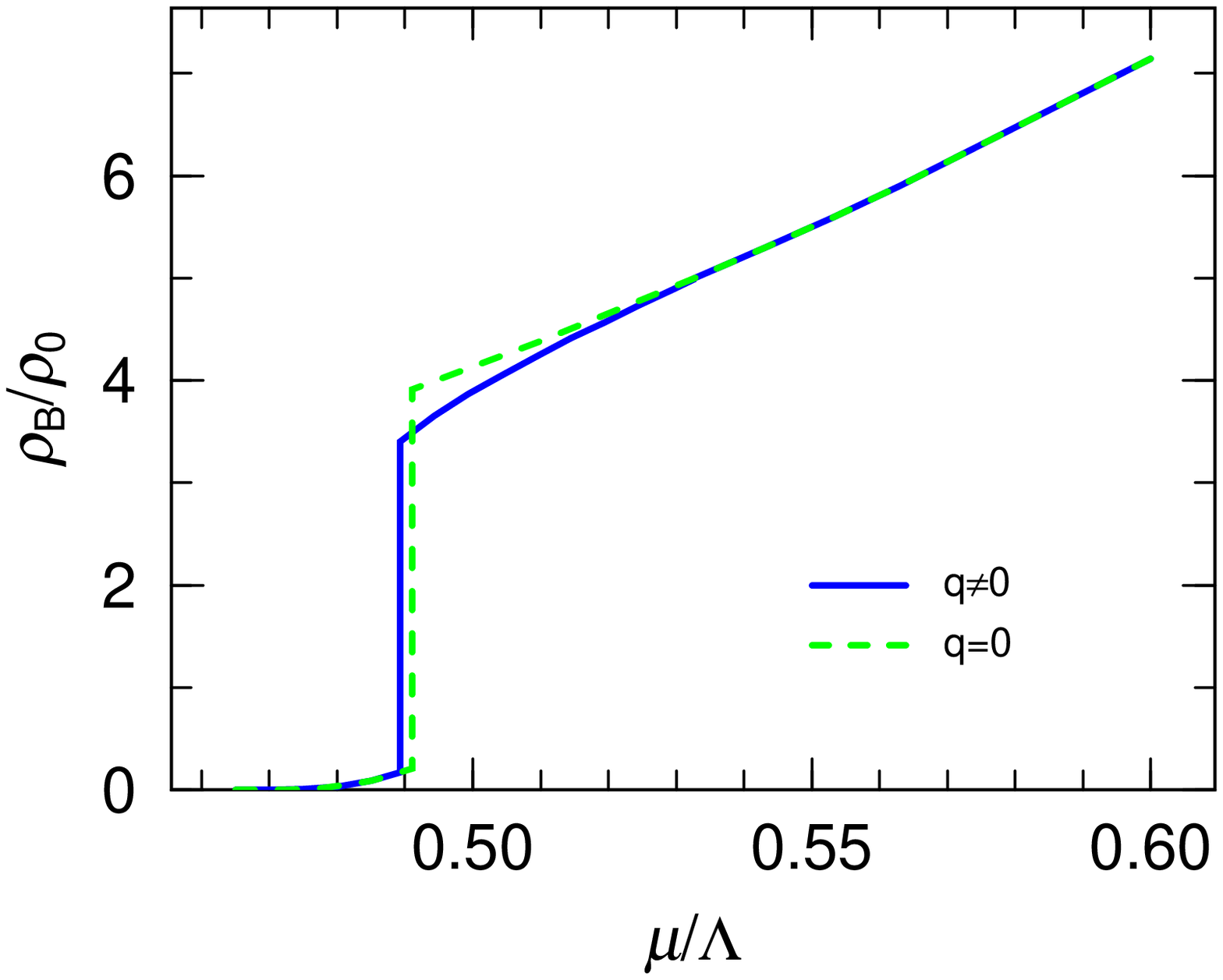}
\end{center}
\vspace*{-0.5cm}
\caption{Baryon number density as a function of $\mu$. 
$\rho_0=0.16 {\rm fm}^{-3}$: the normal nuclear density.
Solid (dashed) line is for the finite $q$ ($q=0$).}
\label{rhoCDW}
\end{figure}

The Fig.~\ref{op1} shows the behaviors of order-parameters $M$ and $q$ 
as functions of $\mu$ at $T=0$, 
where that of $M$ without the density wave is also shown for comparison. 
% 
%------------------------------------------------------
%------------------------------------------------------   
% 
It is found from the figure that the magnitude of $q$ becomes finite 
just before the critical point of the usual chiral transition, 
and DCDW survives at the finite range of $\mu$ ($\mu_{c1}\le\mu\le\mu_{c2}$) 
where the dynamical mass is reduced 
in comparison with that before the transition, 
and decreases with $\mu$. 
On the other hand, 
the wavenumber $q$ increases with $\mu$, 
but its value is smaller than twice the Fermi momentum 
$2k_F$($\simeq 2\mu$ for free quarks) 
due to the higher dimensional effect; 
the nesting of Fermi surfaces is incomplete in the present 3-D system.  
Actually, the ratio of the wavenumber and the Fermi momentum 
(at normal phase $q=M=0$) becomes $q/k_F=1.17-1.47$ 
for the baryon-number densities
$\rho_b/\rho_0=3.62-5.30$ where DCDW develops.  
The baryon-number density is shown in Fig.~\ref{rhoCDW} as a function of $\mu$ 
for the normal and the density wave cases. 
The jumps of the baryon-number density reflects the first-order transition. 
%------------------------------------------------------
%------------------------------------------------------  
In the DCDW phase, the relation $q/2>M$ is retained, 
and the Fermi surface looks like Fig.~\ref{FS1}(b). 

\begin{figure}[h]
\begin{center}
\includegraphics[height=6cm]{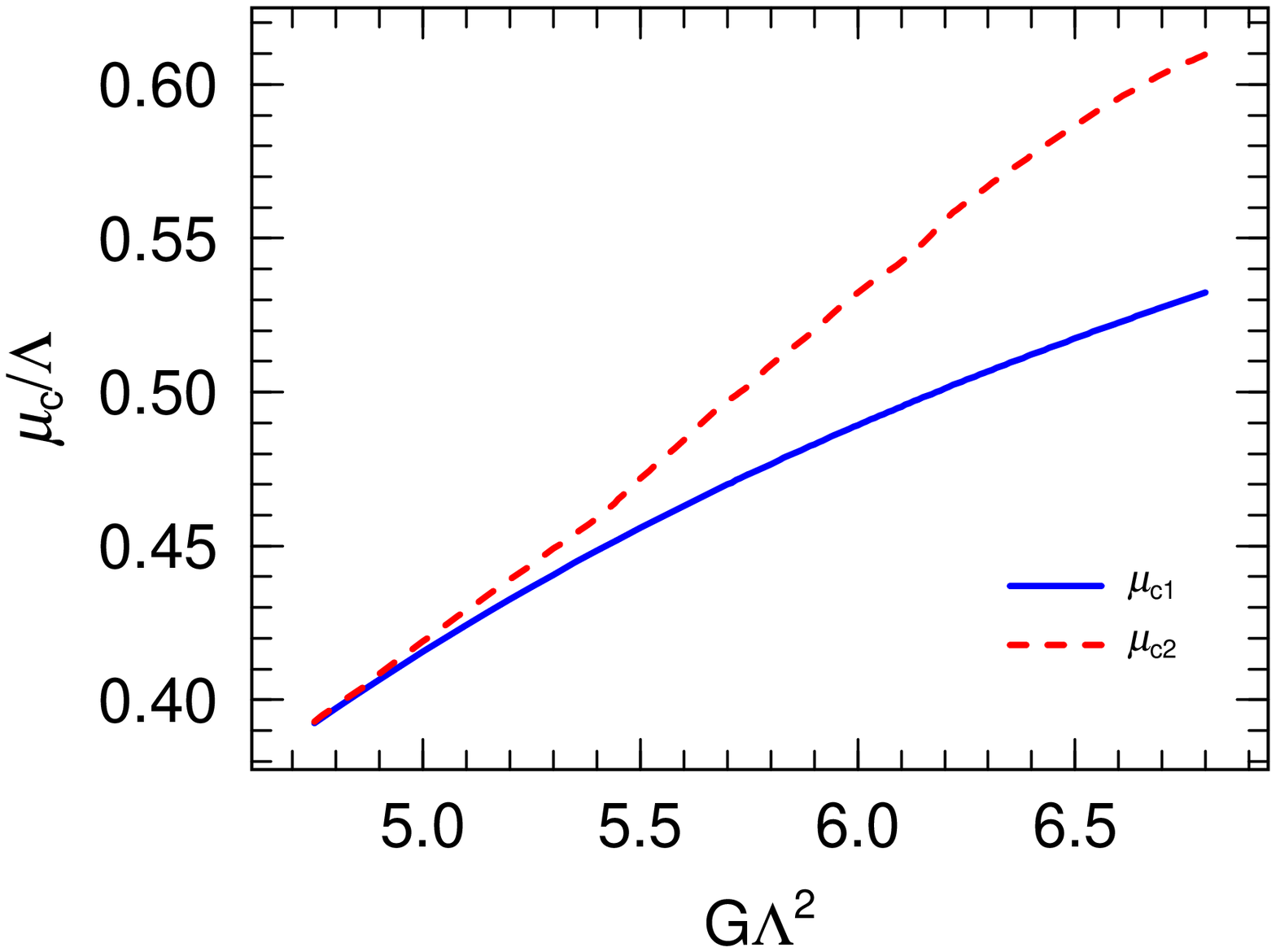}
\end{center}
\caption{Critical chemical potentials $\mu_{c1, c2}$ are plotted as functions of 
the dimensionless coupling $G\Lambda^2$ at $T=0$. 
DCDW phase appears between $\mu_{c1}$ and $\mu_{c2}$. 
The critical coupling at $\mu_{c1}=\mu_{c2}$ 
is estimated to be $G\Lambda^2\simeq 4.72$.}
\label{GvsMuc}
\end{figure}

%%%%%%%%%%%%%%%Newly-added-2%%%%%%%%%%%%%%%%%%%%%%%%%%%%%%%%%
Here we show the coupling-strength dependence of the critical chemical potentials 
$\mu_{c1, c2}$ 
in Fig.~\ref{GvsMuc}, 
including the semi-empirical value, $G\Lambda^2=6.35$ ($\Lambda=660.37$ MeV),  
to reproduce the pion decay constant $f_\pi=93$ MeV 
and the constituent-quark mass $\simeq 330$ MeV, 
one third of the nucleon mass in the vacuum.  
The range of DCDW phase between $\mu_{c1}$ and $\mu_{c2}$ starts to open 
at $G\Lambda^2\simeq 4.72$, 
and broadens with increase of the coupling strength. 
It should be noted that the effective potential (\ref{therm}) can be scaled 
by the cut-off $\Lambda$ 
and thus the dimensionless coupling $G\Lambda^2$ becomes only one parameter 
to determine whether or not the phase transition itself occurs in the present model. 
%%%%%%%%%%%%%%%Newly-added-2%%%%%%%%%%%%%%%%%%%%%%%%%%%%%%%%%

It should be kept in mind that the order of the transitions may depend on 
the parameter choice of $G$ and $\Lambda$, 
and may also on the regularization scheme.

%%%%%%%%%%%%%%%%%%%%%%%%%%%%%%%%%%%%%%%%%%%%%%%%%%%%%%%%%%%%%%%%%%%%%%
\subsection{Magnetic properties} 

Using the eigen spinors in eqs.~(\ref{sp1})-(\ref{sp3}), 
we can calculate various expectation values 
with respect to the DCDW state.  
For a operator, $\mathcal{O}$, which does not depend on the spatial coordinate, 
its expectation value becomes a simple form: 
\beq
\langle \psi^\dagger(r)\mathcal{O} \psi(r) \rangle 
=\idpn \langle \tilde{\psi}^\dagger(p) ~e^{i\tau_3\gamma_5 \bq \cdot \br/2} \mathcal{O} 
e^{-i\tau_3\gamma_5 \bq \cdot \br/2} ~\tilde{\psi}(p) \rangle. 
\eeq   

We can confirm that baryon-number density, $\mathcal{O}=1$, 
is still constant even in the density wave state: 
summation of quasi-particle state in momentum space, 
\begin{eqnarray}
\rho_B=\idpn  \langle \tilde{\psi}^\dagger(p) \tilde{\psi}(p) \rangle=constant.  
\end{eqnarray}

On the other hand, 
the spin expectation value, 
$\mathcal{O}=\gamma_0\gamma_5\gamma_3/2\equiv \Sigma_z$, 
vanishes in each flavor,  
\begin{eqnarray}
\langle \Sigma_z \rangle \equiv 
\frac{1}{2} \idpn  \langle \tilde{\psi}^\dagger(p)  
\gamma_0\gamma_5\gamma_3\tilde{\psi}(p) \rangle=0, 
\end{eqnarray}
because the stationary condition for the wavenumber 
$q$ is proportional to the expectation value: 
\begin{eqnarray}
0=\frac{\partial \Omega_{tot}(q, M)}{\partial q}
\propto \langle \Sigma_z \rangle. 
\label{MZ}
\end{eqnarray}

Here we show an interesting feature of DCDW: 
a spatial modulation of the anomalous magnetic moment.  
The Gordon decomposition of the gauge coupling term gives 
the magnetic interaction with external field $F^{\mu\nu}$ 
in the form,  
$g_L (e^*/2M) (\bar{\psi} \sigma_{\mu\nu} \psi) F^{\mu\nu}$,  
where $g_L$ is a form factor and $e^*$ an effective electric charge. 
The operator of the magnetic moment for the $z$ component is defined by  
$\mathcal{O}=\gamma_0\sigma_{12}$, 
which is not commuted to $\gamma_5$, 
\beq
e^{i\tau_3\gamma_5 \bq \cdot \br/2} \gamma_0\sigma_{12} 
e^{-i\tau_3\gamma_5 \bq \cdot \br/2} 
&=& \gamma_0\sigma_{12} \cos(\bq \cdot \br)
-i\gamma_3 \sin({\bf q}\cdot {\bf r}), 
\eeq
and then its expectation value is given by 
\begin{eqnarray}
\langle \bar{\psi}(r) \sigma_{12} \psi(r) \rangle 
&=&
\langle \gamma_0\sigma_{12}\rangle \cos({\bf q}\cdot {\bf r})
-i\langle \gamma_3 \rangle \sin({\bf q}\cdot {\bf r}), 
\label{Mag01} \\
\mbox{where}~~
\langle \gamma_0\sigma_{12} \rangle
&=& \idpn   
\frac{2M}{\sqrt{M^2+p_z^2}} 
\left[ n_+({\bp}) - n_-({\bp}) \right], 
\label{Mag02} \\
\langle \gamma_3 \rangle
&=& 0. 
\end{eqnarray}
The function $n_\pm(\bp)$ is the momentum distribution 
for the eigen state corresponding to $E_\pm(\bp)$.  
The expectation value $\langle \gamma_0 \sigma_{12} \rangle$ is proportional 
to an asymmetry of the momentum distribution in $n_\pm(\bp)$.  
In DCDW phase, 
the asymmetry becomes finite as shown in Fig.~\ref{FS1}, 
and thus the magnetic moment is spatially modulated with wavenumber $q$.  
The expectation value $\langle \gamma_3 \rangle$ vanishes in each eigen spinor: 
$u^\dagger_\pm \gamma_3 u_\pm=v^\dagger_\pm \gamma_3 v_\pm=0$. 
We also confirmed that the other components of the magnetic moment, $\gamma_0\sigma_{23, 13}$, 
vanished analytically after the integration in momentum space.  

The equation (\ref{Mag01}) shows that the amplitude of the modulated magnetic moment 
depends on the dynamical mass, 
reflecting delay of chiral restoration due to the presence of DCDW.  
The magnetic order of DCDW should have some observable consequence 
of compact stars with quark cores.  
%%%%%%%%%%%%%%%%Newly-added-3%%%%%%%%%%%%%%%%%%%%%%%%%%%%%%%%%%%%
We estimate the magnitude of the amplitude (\ref{Mag02}) at $\rho_B/\rho_0=3\sim4$: 
the tensor expectation value per quark is calculated to be 
$\langle \gamma_0 \sigma_{12} \rangle/\langle \gamma_0 \rangle=0.1\sim0.3$. 
Thus a local magnetic flux $\Phi$ induced by DCDW, 
\begin{eqnarray}
\Phi=
\lk \frac{2}{3}-\frac{1}{3}\rk \frac{e}{2m_q} 
\frac{\langle \gamma_0 \sigma_{12} \rangle}{\langle \gamma_0 \rangle} 3 \rho_B, 
\end{eqnarray}
amounts to $O(10^{16})$ Gauss, 
which is comparable with observed values in magnetars \cite{Mag1}. 
The flux strength on the star surface from the quark core might be smaller, 
since it is given by summation of the quark-magnetic moment (\ref{Mag01})  
modulated rapidly with the wave length $q/2\pi \simeq O(10)$ fm, 
nevertheless, contributions from near the quark-core surface may remain 
without the cancellation. 
%%%%%%%%%%%%%%%%Newly-added-3%%%%%%%%%%%%%%%%%%%%%%%%%%%%%%%%%%%%

%%%%%%%%%%%%%%%%%%%%%%%%%%%%%%%%%%%%%%%%%%%%%%%%%%%%%%%%%%%%%%%%%%%%%%
\subsection{Correlation functions} 
%%%%%%%%%%%%%%%%%%%%%%%%%%%%%%%%%%%%%%%%%%%%%%%%move to correlation fun. 
In this section, 
we consider scalar- and pseudoscalar-correlation functions, $\Pi_{\rm s, sp}(k)$, 
at the chirally restored phase,  
and discuss their relation with DCDW at $T=0$. 
The correlation functions depend on an external four-momentum $k=(k_0, {\bf k})$, 
chemical potential, and also the effective quark mass for a given chemical potential.    
In the static limit $k_0\rightarrow 0$, 
the correlation functions have a physical correspondence 
to the static susceptibility functions for the spin- or charge-density wave \cite{kagoshima}, 
while they have no primal singularity at $|{\bf k}|=2k_F$ reflecting the higher dimensionality. 
Note that the functions have a differential singularity at $k=2k_F$. 
 
In general, a pole of the correlation function implies a second-order phase transition. 
As for the case of DCDW, 
the effective potential analysis in the previous section 
shows the first-order transtion. 
Nevertheless, as shown in Fig.~\ref{cp1}(f),  
the potential barrier from DCDW to the normal phase 
at the critical point $\mu=\mu_{c2}$ is very small, 
and it is expected that the correlation function 
has some imformations of the transition point. 

\begin{figure}
\begin{center}
\includegraphics[height=5.5cm]{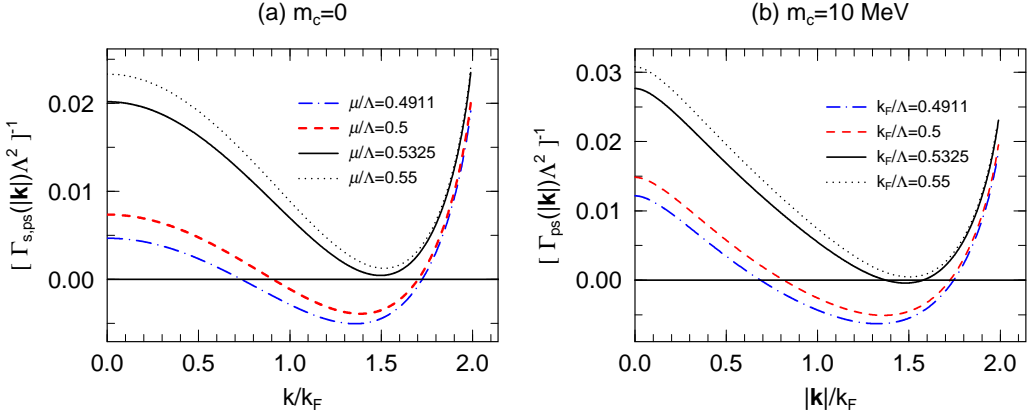}
\end{center}
\caption{Function, $1/\Gamma_{\rm ps}(|\bk|)$, is plotted 
for various values of $\mu$, 
$\mu/\Lambda=0.4911, 0.5, 0.53254(=\mu_{c2})$, and $0.55$, 
for $m_c=0$ (a) and $m_c=10$ MeV (b). 
In the case of $\mu/\Lambda\ge 0.4911$, 
the mass-gap equation has a extremum solution $M=0$. }
\label{corrfn1}
\end{figure}

We evaluate effective interactions, $\Gamma_{\rm s, sp}(k)$,    
within the random phase approximation \cite{nam, kle}, 
which are related to the correlation functions, 
i.e.,  $2G\Pi_{\rm s, sp}(k)=\Gamma_{\rm s, sp}(k)\Pi_{\rm s, sp}^0(k)$:  
\begin{eqnarray}
i\Gamma_{\rm s, ps}(k)
&=& \frac{2Gi}{1-2 G\Pi_{\rm s, ps}^0(k)}, 
\end{eqnarray}  
where $\Pi_{\rm s, ps}^0(k)$ is the polarization function in medium,  
which has form at the static and chiral limits (see Appendix~\ref{corr1}): 
\begin{eqnarray}
\Pi_{\rm s}^0(|\bk|)=\Pi_{\rm ps}^0(|\bk|)&=& 
\frac{N_f N_c }{4\pi^2}(\Lambda^2-2k_F^2)-2N_f N_c i\bk^2I(\bk^2)|_{M\rightarrow0} \nn 
&+&\frac{N_f N_c |\bk|}{4\pi^2}\left[
k_F\log\left( \frac{2k_F+|\bk|}{2k_F-|\bk|} \right) 
+\frac{|\bk|}{2}\log\left\{ \frac{\lk 2k_F+|\bk|\rk \lk 2k_F-|\bk| \rk}{|\bk|^2} \right\} 
\right]. 
\end{eqnarray}  
Inverse of the correlation function in the chiral limit 
corresponds to the coefficient of $M^2$ of the effective potential, 
\begin{eqnarray}
\Omega_{tot}=\Omega_{tot}|_{M\rightarrow0}
+\frac{1}{2}\Gamma_{\rm ps}^{-1}(q)|_{M\rightarrow0} M^2+O(M^4). 
\label{expM}
\end{eqnarray} 
From behaviors of the function $\Gamma_{\rm ps}(|\bk|)^{-1}$ 
shown in Fig.~{\ref{corrfn1}}(a), 
it is found that the function takes the lowest value 
at a finite external momentum ($|\bk| \simeq 1.5 k_F$), 
and thus a finite wavenumber $q$ gives the lower potential energy in eq.~(\ref{expM}).  
Assuming the second-order transition, 
the critical density $\mu_t$ and wavenumber $|\bk_t|$ are determined from 
a following simultaneous equation: 
\begin{eqnarray}
\Gamma_{\rm sp}(|\bk|;\mu)^{-1}=0, ~~\mbox{and}~~ 
\frac{\partial \Gamma_{\rm sp}(|\bk|;\mu)^{-1}}{\partial |\bk|}=0. 
\end{eqnarray}
A numerical calculation in the chiral limit gives 
$\mu_t(=k_F)=0.5320 \Lambda$ and $|\bk_t|=1.498 k_F$, 
which almost coincide with the correct result 
$\mu_{c2}=0.53254 \Lambda$ and $q=1.469 k_F$ 
(where $k_F\equiv\sqrt{\mu_{c2}^2-M^2}$; $M=0.034 \Lambda$)  
from the effective potential analysis. 
A complex structure of the effective potential makes 
a little difference between $|\bk_t|$ and $q$ at $\mu=\mu_{c2}$. 
%%%%%%%%%%%%%%%%%%%%%%%%%%%%%%%%%%%%%%%%%
%------------------------------------------------------
%------------------------------------------------------   

The above argument might also be available 
even for the case of a finite current-quark mass, $m_c\simeq 10$MeV: 
the effective potential for a small current-quark mass ($m_c<<\mu$) is approximated to 
\begin{eqnarray}
\Omega_{tot}\simeq \Omega_{tot}|_{M\rightarrow m_c}+
\frac{1}{2}\Gamma_{\rm ps}^{-1}(q)|_{M\rightarrow m_c} (M-m_c)^2+O[(M-m_c)^4].
\end{eqnarray}
The Fig.~{\ref{corrfn1}}(b) shows that behaviors of the coefficient function 
has little shift from that of the chiral limit, 
and thus suggests a DCDW transition with a small current-quark mass 
as in the chiral limit. 

%%%%%%%%%%%%%%%%%%%%%%%%%%%%%%%%%%%%%%%%%%%%%%%%%%%%%%%%%%%%%%%%%%%%%
\subsection{Phase diagram on the $\mu$-$T$ plane}
%%%%%%%%%%%%%%%%%%%%%%%%%%%%%%%%%%%%%%%%%%%%%%%%%%%%%%%%%%%%%%%%%%%%%
To complete a phase diagram  
we derive the thermodynamic potential at finite temperature 
in the Matsubara formalism. 
The partition function for the mean-field Hamiltonian is given by  
\begin{eqnarray}
&&Z_\beta =
%&&~~=
\int {\it D}\bar{\psi} {\it D}\psi 
\exp \int_0^\beta {\rm d}\tau \int {\rm d}^3r~
 \left\{ \bar{\psi}\left[
i \tilde{\partial} +M \exp\left(i\gamma_5 {\bf q} \cdot {\bf r}\right) 
-\gamma_0 \mu \right]\psi 
-\frac{M^2}{4 G} \right\}, 
%&&~~\tilde{\partial}\equiv -\gamma_0 \partial_\tau+i {\bf \gamma}\nabla. \nn 
\end{eqnarray}
where $\beta=1/T$, 
and $\tilde{\partial}\equiv -\gamma_0 \partial_\tau+i {\bf \gamma}\nabla$. 
Taking the Fourier transform of the spinor with the Matsubara frequency $\omega_n$, 
the partition function becomes 
\begin{eqnarray}
\psi(\br,-i \tau) &=& e^{-i\tau_3\gamma_5\bq\cdot\br/2}~ T \sum_n \idk 
e^{i \omega_n \tau +i \bk \cdot \br} \tilde{\psi}(\bk, n), \nn
Z_\beta &=& 
\prod_{\bk,n,s=\pm} 
\left\{ (i \omega_n+\mu)^2-E_s^2(\bk) \right\}^{N_fN_c}
\times \exp\left\{-\left( \frac{M^2}{4 G}\right)V\beta \right\},  
\end{eqnarray}
where $V$ is volume of the system. 
Thus the thermodynamic potential $\Omega_\beta$ 
per unit volume is obtained,   
\begin{eqnarray}
\Omega_\beta (q,M)
&=&-T \log{Z_\beta(q,M)}/V \nn
%&=&\sum_{\bk,n,n'} T \log 
%\left\{ (i \omega_n+\mu)^2-E_{n'}(\bk)^2 \right\}/V 
%+\frac{M^2}{4 G} \nn
&=&-N_fN_c\idk \sum_{s=\pm} \left\{
T\log{\left[ e^{-\beta\left(E_s(\bk)-\mu \right)} +1 \right] 
      \left[ e^{-\beta\left(E_s(\bk)+\mu \right)} +1 \right]}
+E_s(\bk) \right\} +\frac{M^2}{4 G}\label{tpot} 
\end{eqnarray}
where we have utilized a contour-integral technique 
for the frequency sum to get the final form.

\begin{figure}
\begin{center}
\includegraphics[height=6cm]{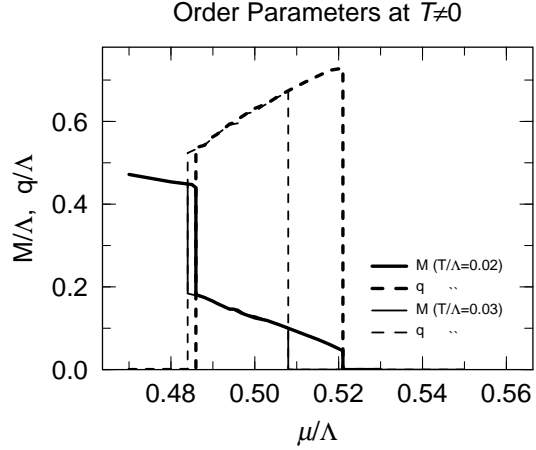}
\end{center}
\caption{The wavenumber $q$ and the dynamical mass $M$ 
are plotted as a function of $\mu$ at finite temperature.
The solid (dashed) line corresponds to $q$ ($M$) 
at $T/\Lambda=0.02$ (thick lines), $0.03$ (thin lines).}
\label{Tops}
\end{figure}

From the absolute minimum of the thermodynamic potential (\ref{tpot}), 
it is found that 
the order parameters at $T\neq0$ behave similarly to those at $T=0$ 
as a function of $\mu$,  
while the chemical-potential range of DCDW,  
$\mu_{c1}(T)\le \mu \le\mu_{c2}(T)$,  
gets smaller as $T$ increases. 
The Fig.~\ref{Tops} shows the order parameters at finite temperature.  
%------------------------------------------------------
%------------------------------------------------------   
The discontinuities of the order parameters reflect the two absolute minima 
at the critical chemical potential $\mu_{c1,2}$, 
and it indicates a first-order transition.  
Thus the region of DCDW in the $\mu$-$T$ phase diagram 
is surrounded by the first-order transition lines.   

\begin{figure}
\begin{center}
\includegraphics[height=6cm]{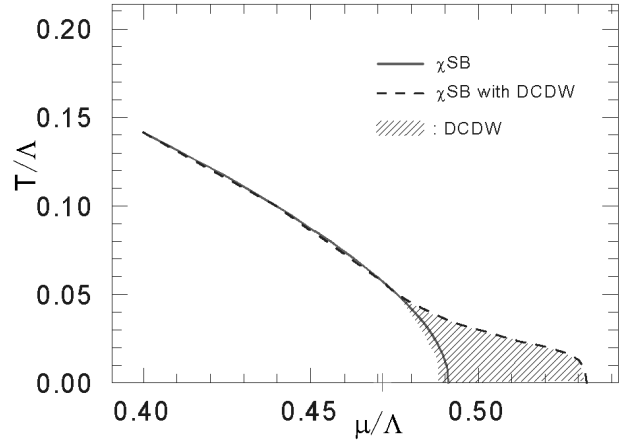}
\end{center}
\caption{A phase diagram obtained 
from the thermodynamic potential eq.~(\ref{tpot}). 
The solid (dashed) line shows the chiral-transition line 
without (with) the DCDW. 
The shaded area shows the DCDW phase. }
\label{pd1}
\end{figure}

We show the resultant phase diagram in Fig.~\ref{pd1}, 
where the usual chiral-transition line is also given for reference. 
%------------------------------------------------------
%------------------------------------------------------   
Comparing phase diagrams with and without $q$,  
we find that the DCDW phase emerges in the area 
(shaded area in Fig.~\ref{pd1})
which lies just outside the boundary of the ordinary chiral transition.
We thus conclude that DCDW is induced by finite-density contributions, 
and has the effect to extend the chiral-condensed phase ($M\neq 0$) 
to a low temperature ($T_c \sim 50$MeV) and high density region. 
The above results suggests that QCD  at finite density involves 
rich and nontrivial phase structures, as well as color superconducting phases. 
%%%%%%%%%%%%%%%%%%%%%%%%%%%%%%%%%%%%%%%%%%%%%%%%%%%%%%%%%%%%%%%%%%%%%%

\section{Summary and outlooks}

We have discussed a possibility of the dual chiral-density wave 
in moderate-density quark matter within the mean-field approximation, 
employing 2-flavor and 3-color NJL model. 
The mechanism of the density wave is quite similar to the spin density wave 
in 3-D electron systems; 
the total-energy gain comes from the Fermi-sea contribution in the deformed spectrum, 
while its amplitude has the different origin  
corresponding to the chiral condensation from the Dirac-sea effect.  

%%%%%%%%%%%%%Newly-revised-2%%%%%%%%%%%%%%%%%%%%%%%%%%%%
In this paper,  
we have considered only the direct channels (Hartree terms) of the interaction. 
If the exchange channels (Fock term) are involved, 
there appear additional interaction channels 
by way of the Fierz transformation \cite{kle}. 
In particular, 
self-energies in axial-vector and tensor channels 
related to a ferromagnetism \cite{nak, mar} might affect the density wave 
through nontrivial correlations among them. 
%%%%%%%%%%%%%Newly-revised-2%%%%%%%%%%%%%%%%%%%%%%%%%%%%%%%%%%%%%
Interactions in the {\it p-p} channels are also obtained 
by the Fierz transformation, 
and their strength is smaller than that of the direct channels 
by the factor of $O(1/N_c)$. 
Because the Cooper instability is independent on the strength of the interaction, 
it is interesting 
to investigate the interplay among the density wave, 
superconductivity \cite{der1,der2,Ohwa}, and the other ordered phases, 
%%%%%%%%%%%%%%second-revised-2%%%%%%%%%%%%%%%%%%%%%%%%%%%%%%%%%%%%
e.g., 
chiral-density waves mixing isospins which may cause a charge-density wave 
due to difference of electric charges of u- and d-quarks, 
%%%%%%%%%%%%%%%%%%%%%%%%%%%%%%%%%%%%%%%%%%%%%%%%%%%%%%%%%%%%%%%%%
as future studies.   

Finally, 
it is worth mentioning about correlation functions 
(fluctuation modes) on the density-wave phase, 
which give the excitation spectrum, 
and are important for the dynamical description of the phase.   
In particular, 
Nambu-Goldstone modes are essential degrees of freedom 
for low-energy phenomena, and may bring some observable consequences, 
%%%%%%%%%%%%%%%%%%second-revised-4%%%%%%%%%%%%%%%%%%%%%%%%%
e.g., slowing down of star cooling through 
enhancement of specific heat due to fluctuations of such low-energy modes. 
%%%%%%%%%%%%%%%%%%%%%%%%%%%%%%%%%%%%%%%%%%%%%%%%%%%%%%%%%%%

%%%%%%%%%%%%%%%%%%%%%%%%Acknowledgement%%%%%%%%%%%%%%%%%%%%%%%%%%%%%%%%%%%%%%
\begin{center}
{\bf Acknowledgements}
\end{center}

Authors thank T. Maruyama, K. Nawa, and H. Yabu for discussions and comments. 
The present research is
partially supported by the Grant-in-Aid for the 21st century COE 
``{\it Center for Diversity and Universality in Physics}'' 
of the Ministry of Education, Culture, Sports, Science and Technology, 
and by the Japanese Grant-in-Aid for Scientific
Research Fund of the Ministry of Education, Culture, Sports, Science and
Technology (11640272, 13640282).
%%%%%%%%%%%%%%%%%%%%%%%%%%%%%%%%%%%%%%%%%%%%%%%%%%%%%%%%%%%%%%%%%%%%%%%%%%%%%%

\appendix
%%%%%%%%%%%%%%%%%%%%%%%%%%%%%%%%%%%%%%%%%%%%%%%%%%%%%%%%%%%%
\section{Regularization of $\Omega_{\rm vac}$ \label{PTR1} }

We regularize the Dirac-sea contributions to the potential, $\Omega_{\rm vac}$, 
by applying the schwinger's proper-time method. 
$\Omega_{\rm vac}$ can be described 
in the form of the one-loop order contribution, 
\begin{eqnarray}
\Delta\Omega=\Omega_{\rm vav} (q,M) - \Omega_{\rm N} 
%&=&-\int_C \frac{{\rm d}^4k}{i(2\pi)^4}\log \frac{\det D_A}{\det D_N} \\
&=&-\int_C \frac{{\rm d}^4k}{i(2\pi)^4}\sum_{s=\pm} \log \frac{D_s}{D_N} \\
\mbox{with} &&
D_\pm=k_0^2-E_\pm^2({\bf k}) ~~\mbox{and}~~
D_N=k_0^2-{\bf k}^2-m^2, 
\label{tpot0}
\end{eqnarray}
where $\Omega_{\rm N}$ is the normal vacuum contribution. 
Using the identity for $G\in {\bf R}$ 
\begin{eqnarray}
(G+i\eta)^{-1}=-i\int_0^\infty {\rm d}s ~e^{i(G+i\eta)s}, 
\end{eqnarray}
we find 
\begin{eqnarray}
\log \frac{D_\pm+i\eta}{D_N+i\eta}=-\int_0^\infty \frac{{\rm d}s}{s} 
\left( e^{i(D_\pm+i\eta) s}-e^{i(D_{\rm N}+i\eta) s} \right).
\end{eqnarray}

By way of the Wick rotation 
which is done simultaneously for $k_0$ integration of 
$\Omega_{\rm vac}$ and $\Omega_{\rm N}$, 
\begin{eqnarray}
\Delta\Omega_\pm
&=&\int_C \frac{{\rm d}^4k}{i(2\pi)^4}
\int_0^\infty \frac{{\rm d}s}{s} 
\left( e^{iD_\pm s}-e^{iD_0 s} \right) \nn
&=&\int_{-\infty}^\infty \frac{{\rm d}^4k_E}{(2\pi)^4}
\int_0^\infty \frac{{\rm d}s}{s} 
\left[ \exp i\{-k_E^2-k_t^2 -(\sqrt{k_z^2+M^2}\pm q/2)^2\} s \right. \nn
&&~~~~~~~~~~~~~~~~~~~~~~~~~~~~~~~~~-\left. \exp i\{-k_E^2-k_t^2-k_z^2-m^2\} s \right] \nn 
&=&\int_{-\infty}^\infty \frac{{\rm d}^4k_E}{(2\pi)^4}
\int_0^\infty \frac{{\rm d}\tau}{\tau} 
\left[ \exp \{-k_E^2-k_t^2 -(\sqrt{k_z^2+M^2}\pm q/2)^2\} \tau \right. \nn
&&~~~~~~~~~~~~~~~~~~~~~~~~~~~~~~~~~-\left. \exp \{-k_E^2-k_t^2-k_z^2-m^2\} \tau \right] \nn 
&=&\frac{1}{8\pi^{3/2}}\int_0^\infty \frac{{\rm d}k_z}{2 \pi} 
\int_0^\infty \frac{{\rm d}\tau}{\tau^{5/2}} 
\left[
\exp \{-(\sqrt{k_z^2+M^2}\pm q/2)^2 \tau \}
-\exp \{-(k_z^2+m^2) \tau \} \right] \nn
\end{eqnarray}

The above integration of $\tau$ has singular at $\tau=0$ and thus not well defined.
The proper-time regularization is to replace the lower limit of $\tau$ 
by the cutoff $1/\Lambda^2$, 
\begin{eqnarray}
\int_0^\infty {\rm d}\tau \rightarrow  \int_{1/\Lambda^2}^\infty {\rm d}\tau, 
\end{eqnarray}
the $\Lambda$ corresponds to a momentum cutoff.

Eventually we obtain the regularized potential from the Dirac sea,
\begin{eqnarray}
\Delta\Omega &=&\Delta\Omega_+ +\Delta\Omega_- \nn
&=&\frac{1}{8\pi^{3/2}}\int_0^\infty \frac{{\rm d}k_z}{2 \pi} 
\int_{1/\Lambda^2}^\infty \frac{{\rm d}\tau}{\tau^{5/2}} 
\left[
\exp \{-(\sqrt{k_z^2+M^2}+ q/2)^2 \tau \}+\exp \{-(\sqrt{k_z^2+M^2}-q/2)^2 \tau \} \right. \nn
&&~~~~~~~~~~~~~~~~~~~~~~~~~~~~~~~~~~~~~~~~~~~
~\left.-2 \exp \{-(k_z^2+m^2) \tau \} \right]. 
\end{eqnarray}

The normal vacuum contribution $\Omega_N$ has an explicit form,  
\begin{eqnarray}
\Omega_N&=& N_c N_f \frac{1}{8 \pi^2} \int_{1/\Lambda^2}^\infty \frac{{\rm d}\tau}{\tau^3} 
e^{-m^2 \tau} \nn
&=& N_c N_f \frac{\Lambda^4}{16 \pi^2} 
\left\{ (-\tilde{m}^2+1)e^{-\tilde{m}^2}+ \tilde{m}^4 \Gamma(0,\tilde{m}^2) \right\}, 
\end{eqnarray}
where $\tilde{m}=m/\Lambda$ and 
$\Gamma(a,z)=\int_z^\infty {\rm d} \tau \tau^{a-1} \exp(-\tau)$ 
is the incomplete gamma function.

%%%%%%%%%%%%%%%%%%%%%%%%%%%%%%%%%%%%%%%%%%%%%%%%%%%%%%%%%%%%%%%%%%%%%%%%%%%%%%%%%%%%
\section{Fock exchange contributions \label{fock} }

We briefly examine how the Fock exchange terms of the NJL interaction affect DCDW. 
After the Fierz transformation, 
one can find the exchange interaction terms, 
discarding color-octet contributions \cite{kle}: 
\begin{eqnarray}
\mathcal{F}\ldk \lk \bar{\psi}\psi \rk^2
+ \lk \bar{\psi} i\gamma_5 {\bf \tau} \psi \rk^2 \rdk
&=& 
\frac{1}{8N_c}\ldk 
 2\lk \bar{\psi}\psi \rk^2
+2\lk \bar{\psi} i\gamma_5 {\bf \tau} \psi \rk^2 
-2\lk \bar{\psi}{\bf \tau} \psi \rk^2
-2\lk \bar{\psi} i\gamma_5 \psi \rk^2  \right. \nn 
&&- \left. 4\lk \bar{\psi} \gamma_\mu \psi \rk^2
-4\lk \bar{\psi} i\gamma_5 \gamma_\mu \psi \rk^2
+\lk \bar{\psi} \sigma_{\mu\nu} \psi \rk^2
-\lk \bar{\psi} \sigma_{\mu\nu} {\bf \tau} \psi \rk^2
\rdk
\end{eqnarray}
The overall factor $1/8N_c$ indicates that the exchange terms are 
less relevant in comparison with the direct terms. 

The first two terms come to be added to the Hartree terms, 
and contribute to DCDW 
through changing the effective interaction in these channels by a factor. 

The third and fourth terms have opposite sign of interactions, 
and thereby they can not gain condensation energies. 

The fifth one has an expectation value in its temporal term, 
which corresponds to the baryon number density, 
$\langle \psi^\dagger \psi \rangle \neq 0$, 
and is renormalized into the chemical potential \cite{hatsu, asa}. 
Thus it might not change DCDW qualitatively, 
except for changing value of a bare chemical potential 
for a fixed baryon number density. 
The other spatial terms vanish self-consistently 
in the present formalism \cite{nak}. 

The sixth term, the axial-vector interaction, seems to contribute DCDW 
because the Weinberg transformation leaves it unchanged, 
but scalar and pseudo-scalar terms for DCDW 
changed to an another axial-vector mean-field with isospin $\tau_3$. 
However, it is found that this term does not affect DCDW by itself: 
the Hartree-Fock free-energy with both the axial-vector mean-field $V_A$ 
and DCDW (the wavenumber $q$) is given, 
after the Weinberg transformation, by  
\begin{eqnarray}
{\cal H}_{HF} 
&=& 
\int \bar{\psi} \ldk \bp^2 +M^2 
+\gamma_3\gamma_5\lk V_A +\tau_3 q/2 \rk \rdk \psi
%+\int \bar{\psi}_d \ldk \bp^2 +M^2 +\gamma_3\gamma_5\lk V_A-q/2 \rk \rdk \psi_d 
+\frac{M^2}{4G_S}+\frac{V_A^2}{4G_A}, \nn 
&=& 
\int \bar{\psi}_u \ldk \bp^2 +M^2 
+\gamma_3\gamma_5 A \rdk \psi_u 
+\int \bar{\psi}_d \ldk \bp^2 +M^2 
+\gamma_3\gamma_5 B \rdk \psi_d +\frac{M^2}{4G_S}+\frac{(A+B)^2}{16G_A},  
\label{effl} \\
A&=&V_A+q/2,~\mbox{and}~B=V_A-q/2, \nonumber 
\end{eqnarray}
where $\psi_{u,d}$ are spinors of u,d-quark, and 
$G_{S,A}$ are reduced coupling constants in scalar and axial-vector channels. 
The equation (\ref{effl}) shows that 
the minimum of the effective potential is always given by $A+B=2V_A=0$ 
because the first two terms have the same structure in their energy spectrum 
with respect to $A, B$ independently of the sign of $A, B$,
and implies, therefore, that the expectation value of the axial-vector channel vanishes. 

The last two terms, the tensor interactions, 
might give a non trivial contribution to DCDW 
through an feedback effect of the nonzero expectation value of the tensor operator, 
$\langle \gamma_0 \sigma_{12} \rangle$, see eq.~(\ref{Mag01}), 
although their interaction strengths are small by the factor $1/8N_c$. 
This problem is left for a future study.

%%%%%%%%%%%%%%%%%%%%%%%%%%%%%%%%%%%%%%%%%%%%%%%%%%%%%%%%%%%%%%%%%%%%%%%%%%%%%%%%%%%%
\section{Scalar and pseudo-scalar scattering amplitudes \label{corr1} } 
Following Nambu \cite{nam}, 
we consider the quark-quark scattering matrix generated by the chain diagram 
in the pseudo-scalar channel. 
Then the polarization function, $-i\Pi_{\rm ps}^0$, is given by 
\begin{eqnarray} 
-i\Pi_{\rm ps}^0(k^2)=-\int\frac{{\rm d}^4p}{(2\pi)^4}{\rm tr}
\left[i\gamma_5\tau_3 iS(p+\frac{1}{2}k)i\gamma_5\tau_3iS(p-\frac{1}{2}k)\right],
\end{eqnarray} 
with the quark propagator in medium,  
\begin{eqnarray}
S(p)&=&\frac{1}{\sla{p}-m}+i\frac{\pi}{E_p} (\sla{p}+m) \theta(\mu-E_p) \delta(p_0-E_p) \nonumber  \\
&\equiv & (\sla{p}+m) \tilde{S}(p)= 
(\sla{p}+m) \left[ \tilde{S}_F(p) +\tilde{S}_D(p) \right]. 
\end{eqnarray}
Then the scattering matrix $M_{33}^{\rm ps}$ can be written in the form,
\begin{eqnarray}
iM_{33}^{\rm ps}(k^2)=(i\gamma_5)\tau_3\left[\frac{2iG}{1-2G\Pi^0_{\rm ps}(k^2)}\right](i\gamma_5)\tau_3.
\end{eqnarray}

There are three kinds of contributions to $\Pi_{\rm ps}^0(k^2)$,
\begin{eqnarray}
\Pi_{\rm ps}^0(k^2)=\Pi_{\rm ps}^{FF}(k^2)+\Pi_{\rm ps}^{DD}(k^2)
+2\Pi_{\rm ps}^{DF}(k^2)
\end{eqnarray}
with 
\begin{eqnarray}
-i\Pi_{\rm ps}^{ij}(k^2)=-4N_fN_c\int\frac{{\rm d}^4p}{(2\pi)^4}
\left(-p^2+m^2+\frac{1}{4}k^2\right){\tilde S}_i\left(p+\frac{1}{2}k\right)
{\tilde S}_j\left(p-\frac{1}{2}k\right).
\end{eqnarray}
First, we consider the vacuum contribution, 
\begin{eqnarray}
-i\Pi_{\rm ps}^{FF}(k^2)&=&2N_fN_c\int\frac{{\rm d}^4p}{(2\pi)^4}
\left[\frac{1}{(p+\frac{1}{2}k)^2-m^2}+\frac{1}{(p-\frac{1}{2}k)^2-m^2}\right]
\nonumber\\
&&-2N_fN_ck^2\int\frac{{\rm d}^4p}{(2\pi)^4}\frac{1}{[(p+\frac{1}{2}k)^2-m^2]
[(p-\frac{1}{2}k)^2-m^2]}.
\end{eqnarray}
The integrals in the first term is easily evaluated to get 
\begin{eqnarray}
\int\frac{{\rm d}^4p}{(2\pi)^4}\frac{1}{(p\pm\frac{1}{2}k)^2-m^2}
=-\frac{i}{16\pi^2}\int\frac{{\rm d}\tau}{\tau^2}e^{-m^2\tau},
\end{eqnarray}
in the proper-time representation. 
The integral in the second term is denoted by $I(k^2)$,
\begin{eqnarray}
I(k^2)=\int\frac{{\rm d}^4p}{(2\pi)^4}\frac{1}{[(p+\frac{1}{2}k)^2-m^2]
[(p-\frac{1}{2}k)^2-m^2]}.
\end{eqnarray}
Using the Feynman's trick,
\begin{eqnarray}
I(k^2)&=&\int_0^1{\rm d}t\int\frac{{\rm d}^4p}{(2\pi)^4}\frac{1}{[p^2-m^2+(k^2+2pk)t]^2}
\nonumber\\
&=&\int_0^1{\rm d}t\int\frac{{\rm d}^4p}{(2\pi)^4}\frac{\partial}{\partial m^2}
\frac{1}{p^2-m^2+k^2t(1-t)},
\end{eqnarray}
and introducing the proper-time $\tau$, we find
\begin{eqnarray}
I(k^2)=\frac{i}{16\pi^2}\int_0^1 dt\int_0^\infty\frac{{\rm d}\tau}{\tau}
e^{-(m^2-k^2t(1-t))\tau}. 
\end{eqnarray}

The vacuum contribution is summarized as follows: 
\begin{eqnarray}
-i\Pi_{\rm
ps}^{FF}(k^2)=-\frac{iN_fN_c}{4\pi^2}\int\frac{{\rm d}\tau}{\tau^2}e^{-m^2\tau}
+\frac{iN_fN_c}{8\pi^2}k^2\int_0^1 {\rm d}t\int_0^\infty\frac{{\rm d}\tau}{\tau}
e^{-(m^2-k^2t(1-t))\tau}.
\end{eqnarray}

Secondly, let us consider $\Pi_{\rm ps}^{DF}(k^2)$,
\begin{eqnarray}
-i\Pi_{\rm
ps}^{DF}(k^2)&=&-4N_fN_c\int\frac{{\rm d}^3p}{(2\pi)^4}\left[-\left(\frac{1}{2}k_0+E_{{\bf
p}-\frac{1}{2}{\bf k}}\right)^2+{\bf p}^2+m^2+\frac{1}{4}k^2\right]\nonumber\\
&\times&\frac{1}{\left(k_0+E_{{\bf
p}-\frac{1}{2}{\bf k}}\right)^2-\left({\bf
p}+\frac{1}{2}{\bf k}\right)^2-m^2}\frac{i\pi}{E_{{\bf
p}-\frac{1}{2}{\bf k}}}\theta(\mu-E_{{\bf p}-\frac{1}{2}{\bf k}}).
\end{eqnarray}
Taking the static limit $k_0 \rightarrow 0$, 
we have
\begin{eqnarray}
-i\Pi_{\rm ps}^{DF}(|\bf k|)&\rightarrow&
2N_fN_ci\int\frac{{\rm d}^3p}{(2\pi)^3}\frac{{\bf p}\cdot{\bf k}}{2{\bf
p}\cdot{\bf k}+{\bf k}^2}\frac{1}{E_p}\theta(\mu-E_p)\nonumber\\
&=&2N_fN_ci\int_{-1}^1{\rm d}x\int\frac{p^2{\rm d}p}{(2\pi)^2}
\left[1-\frac{|\bf k|}{2px+|\bf k|}\right]\frac{1}{E_p}\theta(\mu-E_p)\nonumber\\
&=&i\frac{N_fN_c}{4\pi^2}\left[p_F\mu-m^2 
\ln\left(\frac{p_F+\mu}{m}\right)
-\frac{|\bf k|}{2}\int_0^{p_F}p{\rm d}p\ln\left|\frac{2p+|\bf k|}{2p-|\bf k|}\right|
\frac{1}{E_p}\right] \nonumber\\
&&+\frac{N_fN_c |\bf k|}{8\pi}\left[\mu-\left(m^2+\frac{|\bf k|^2}{4}\right)^{1/2}\right].
\end{eqnarray}
The integral over the Fermi sea can be analytically performed, 
but we do not write it down here because of its complexity. 

Finally, we calculate $\Pi_{\rm ps}^{DD}(k^2)$,
\begin{eqnarray}
-i\Pi_{\rm
ps}^{DD}(k^2)&=&-\int\frac{{\rm d}^4p}{(2\pi)^4}4N_fN_c\left(-p^2+m^2+\frac{1}{4}k^2\right){\tilde
S_D\left(p+\frac{1}{2}k\right)}{\tilde
S_D\left(p-\frac{1}{2}k\right)} \nonumber\\
&=&-\int\frac{{\rm d}^3p}{(2\pi)^4}4N_fN_c\left[-\left(\frac{1}{2}k_0+E_p\right)^2
+\left({\bf p}+\frac{1}{2}{\bf k}\right)^2+m^2+\frac{1}{4}k^2\right]\nonumber\\
&\times&\frac{i\pi}{E_{{\bf p}+{\bf k}}}
\theta(\mu-E_{{\bf p}+{\bf k}})\delta(k_0+E_{\bf p}-E_{{\bf p}+{\bf k}})\theta(\mu-E_{\bf p}).
\end{eqnarray}
In the static limit $k_0\rightarrow 0$, 
\begin{eqnarray}
-i\Pi_{\rm ps}^{DD}(|\bk|)\rightarrow 
-N_fN_c\frac{|\bk|}{4\pi}\left[\mu-\left(m^2+\frac{\bk^2}{4}\right)^{1/2}\right],
\end{eqnarray}
which exactly cancels the imaginary part arising from $\Pi_{\rm ps}^{DF}(|\bk|)$. 

Collecting them together, 
we have the denominator of the scattering amplitude in the static limit, 
\begin{eqnarray}
1-2G\Pi_{\rm ps}^0(|\bk|)&=&
1-2G(\Pi_{\rm ps}^{FF}(|\bk|)+2\Pi_{\rm ps}^{FD}(|\bk|)+\Pi_{\rm ps}^{DD}(|\bk|))\nonumber\\
&=&
1-2G\frac{N_fN_c}{4\pi^2}\int\frac{{\rm d}\tau}{\tau^2}e^{-m^2\tau}-4GiN_fN_c\bk^2I(\bk^2)\nonumber\\
&+&G\frac{N_fN_c}{\pi^2}\left[p_F\mu-m^2
\ln\left(\frac{p_F+\mu}{m}\right)-\frac{|\bk|}{2}\int_0^{p_F}p{\rm d}p 
\ln\left|\frac{2p+|\bk|}{2p-|\bk|}\right|\frac{1}{E_p}\right]. 
\end{eqnarray}
On the other hand, the gap equation in this case reads
\begin{eqnarray}
m=m_c+\frac{1}{2\pi^2}GN_fN_cm\int\frac{{\rm d}\tau}{\tau^2}e^{-m^2\tau}
-G\frac{N_fN_c}{\pi^2}m\left[p_F\mu-m^2\ln\left(\frac{p_F+\mu}{m}\right)\right]. 
\end{eqnarray}
Thus, we find at a stationary point (at a solution of the gap equation), 
\begin{eqnarray}
1-2G\Pi_{\rm ps}^0(|\bk|)&=&\frac{m_c}{m}-4GiN_fN_c\bk^2I(\bk^2)
-G\frac{N_fN_c}{2\pi^2}|\bk|\int_0^{p_F}p{\rm d}p\ln\left|\frac{2p+|\bk|}{2p-|\bk|}\right|\frac{1}{E_p}.  
%\nonumber\\
%&=&-4GiN_fN_c(k^2I(-k^2)+m^{*2}_\pi I(m^{*2}_\pi))
%-G\frac{N_fN_c}{2\pi^2}k\int_0^{p_F}p{\rm d}p\ln\left|\frac{2p+k}{2p-k}\right|\frac{1}{E_p}.
\end{eqnarray}

In the similar way, the scattering amplitude of scalar channel is given by 
\begin{eqnarray}
1-2G\Pi_{\rm s}^0(|\bk|)
&=&
1-2G\frac{N_fN_c}{4\pi^2}\int\frac{{\rm d}\tau}{\tau^2}e^{-m^2\tau}-4GiN_fN_c(\bk^2+4m^2)I(\bk^2) \nonumber\\
&+&G\frac{N_fN_c}{\pi^2}\left[p_F\mu-m^2
\ln\left(\frac{p_F+\mu}{m}\right)-\frac{\bk^2+2m^2}{2 |\bk|}\int_0^{p_F}p{\rm d}p 
\ln\left|\frac{2p+|\bk|}{2p-|\bk|}\right|\frac{1}{E_p}\right] \nn 
&+& 2GN_fN_ci \frac{m^2}{2\pi |\bk|} \left[ \mu-\sqrt{m^2+\bk^2/4}\right]
\theta\left(2k_F-|\bk| \right). 
\end{eqnarray}

%%%%%%%%%%%%%%%%%%%%%%%%%%%%%%%%%%%%%%%%%%%%%%%%%%%%

\end{document}